\documentclass[12pt,14paper]{article}
\usepackage[english]{babel}
\usepackage[latin1]{inputenc}
\usepackage[intlimits]{amsmath}
\usepackage{amssymb}
\usepackage{geometry}
\usepackage{verbatim}

\begin{document}

\title{\vspace{-2cm} 
{\normalsize
\vspace{-.4cm}\flushright TUM-HEP 934/14\\
\vspace{-.4cm}\flushright FLAVOUR(267104)-ERC-69\\}
\vspace{0.6cm} 
\bf Radiative Generation of \\Quark Masses and Mixing Angles \\in the Two Higgs Doublet Model \\[8mm]}

\author{Alejandro Ibarra$^1$  and Ana Solaguren-Beascoa$^{1,2}$ \\[2mm]
{\normalsize$^1$ \it Physik-Department T30d, Technische Universit\"at M\"unchen,}\\[-0.05cm]
{\normalsize \it James-Franck-Stra\ss{}e, 85748 Garching, Germany}\\[-0.05cm]
{\normalsize $^2$\it Max-Planck-Institut f\"ur Physik (Werner-Heisenberg-Institut),}\\ [-0.05cm]
{\normalsize \it F\"ohringer Ring 6, 80805 M\"unchen, Germany}
}

\maketitle

\begin{abstract}
We present a framework to generate the quark mass hierarchies and mixing angles by extending the Standard Model with one extra Higgs doublet. The charm and strange quark masses are generated by small quantum effects, thus explaining the hierarchy between the second and third generation quark masses. All the mixing angles are also generated by small quantum effects: the Cabibbo angle is generated at zero-th order in perturbation theory, while the remaining off-diagonal entries of the Cabibbo-Kobayashi-Maskawa matrix are generated at first order, hence explaining the observed hierarchy $|V_{ub}|,|V_{cb}|\ll |V_{us}|$. The values of the radiatively generated parameters depend only logarithmically on the heavy Higgs mass, therefore this framework can be reconciled with the stringent limits on flavor violation by postulating a sufficiently large new physics scale.
\end{abstract}

\section{Introduction}

The quark masses and mixing angles are fundamental parameters in the Standard Model of Particle Physics which must be determined experimentally.  While it is generically expected that dimensionless parameters of the Lagrangian should be either ${\cal O}(1)$ or zero, experiments have revealed hierarchies among the masses of quarks of different generations as well as hierarchies among the quark mixing angles, suggesting the existence of an underlying mechanism generating this structure. 

Several ideas have been discussed in the literature to explain the observed pattern of quark masses and mixing angles. A very popular approach consists in postulating the existence of a ``horizontal'' $U(1)$ symmetry, under which the left- and right-handed quarks of different generations transform differently, and which is assumed to be spontaneously broken at an energy below a certain cut-off. The masses and mixing angles then arise as powers of the small ratio of the $U(1)$ symmetry breaking scale over the cut-off scale~\cite{Froggatt:1978nt}. This approach has been generalized to non-Abelian symmetries, {\it e.g} in~\cite{Barbieri:1998em,King:2001uz} or to discrete symmetries, {\it e.g} in \cite{Ma:2002yp}. A second approach consists in postulating tree level masses for the heavier generation quarks, while the lighter generations acquire masses by quantum effects, thus naturally explaining the observed hierarchy in the quark masses of different generations. Early attempts to radiatively generate fermion masses were presented in~\cite{Weinberg:1972ws,Georgi:1972hy}, based on a gauge group $SU(3)_L\times SU(3)_R$ with the leptons $e^-$, $\nu$ and $\mu^+$ forming a triplet. Since then, many authors have constructed radiative mass models by extending (without horizontal symmetries) the gauge sector, {\it e.g} in  \cite{Mohapatra:1974wk,Balakrishna:1987qd,Balakrishna:1988ks}, or by introducing supersymmetry, {\it e.g.} in \cite{ArkaniHamed:1996zw,Borzumati:1999sp}.

In this letter we will present a mechanism to generate quark mass hierarchies and  mixing angles in the framework of the general two Higgs doublet model. No new fermions nor new symmetries will be introduced.\footnote{A similar approach was pursued in \cite{Ibarra:2011gn,Grimus:1999wm} to generate a mild neutrino mass hierarchy. }  As is well known, this model generically leads to too large flavor violation, hence it is common to impose a discrete symmetry forbidding the simultaneous coupling of two Higgs bosons to the same fermion~\cite{Glashow:1976nt}. However, the flavour violating effects can also be suppressed if the new physics arises at a sufficiently large energy scale. We will show that in this scheme the radiatively generated quark masses are only mildly dependent on the scale of new physics and therefore the same conclusions remain valid even in the decoupling limit. 

\section{Flavor structures in the 2HDM}

The flavor dependent part of the general two Higgs doublet model has the following Lagrangian~\cite{Branco:2011iw}:
\begin{align}\label{eq:YukawaSM}
-{\cal L}^{\rm Yuk}
& = 
 (Y_u^{(a)})_{ij} \bar q_{Li} u_{Rj} \tilde \Phi_a+
(Y_d^{(a)})_{ij} \bar q_{Li} d_{Rj} \Phi_a+{\rm h.c.}\;,
\end{align}
where $i,j=1,2,3$  are flavor indices, $a=1,2$ is a Higgs index and $\tilde{\Phi}_a=i\tau_2\Phi^*_a$. It will be convenient in what follows to work in the Higgs basis where one of the Higgs fields, say $\Phi_2$, does not acquire a vacuum expectation value. Therefore $\langle \Phi^0_1 \rangle =v/\sqrt{2}$, with $v=246$ GeV,  and $\langle \Phi_2^0\rangle=0$. In this basis, then, the Yukawa matrices $Y_{u,d}^{(1)}$ are proportional to the fermion mass matrices.

We will assume, in view of the large mass hierarchy between quarks of different generations, that all the Yukawa matrices have rank 1 at tree level. It can be checked that, by means of a basis transformation of the quark fields, the tree level Yukawa couplings to the Higgs $\Phi_1$ can be written in the form:
\begin{align}
Y_u^{(1)}|_{\rm tree}=
\begin{pmatrix} 
0 & 0& 0 \\ 0 & 0 & 0 \\  0 & 0 & y_u^{(1)}
\end{pmatrix}\;,
\hspace{.2cm}
Y_d^{(1)}|_{\rm tree}=
\begin{pmatrix} 
0 & 0& 0 \\ 0 & 0 & \epsilon y_d^{(1)}\\  0 & 0 & y_d^{(1)}
\end{pmatrix}\;,
\end{align}
which lead to 
\begin{align}
&m_t^{\rm tree}= y_u^{(1)} v/\sqrt{2}\;,&m_c^{\rm tree}=m_u^{\rm tree}=0\;,\nonumber\\
&m_b^{\rm tree}= y_d^{(1)}\sqrt{1+\epsilon^2} v/\sqrt{2}\;,& m_s^{\rm tree}=m_d^{\rm tree}=0\;.
\end{align}
Besides, the elements of the Cabibbo-Kobayashi-Maskawa (CKM) matrix fulfill $\left| V_{ub} \right|^2+\left| V_{cb} \right|^2=\epsilon^2$ , while $V_{us}$ is not defined, since any rotation between the left-handed quarks of the first and second generation leaves the Lagrangian invariant. Experimentally $\left| V_{ub} \right|^2+\left| V_{cb} \right|^2\ll 1$, hence we will assume in what follows that $\epsilon=0$.

On the other hand, the Yukawa couplings to the Higgs $\Phi_2$ must take the most general form of a rank-1 matrix, namely:
\begin{align}
Y_u^{(2)}|_{\rm tree}=
U_L^\dagger
\begin{pmatrix} 
0 & 0& 0 \\ 0 & 0 & 0 \\  0 & 0 & y_u^{(2)}
\end{pmatrix}
U_R\;, \nonumber \\
Y_d^{(2)}|_{\rm tree}=
D_L^\dagger
\begin{pmatrix} 
0 & 0& 0 \\ 0 & 0 & 0\\  0 & 0 & y_d^{(2)}
\end{pmatrix}
D_R\;,
\end{align}
where $U_{L,R}$, $D_{L,R}$ are $3\times3$ unitary matrices. The Yukawa matrix elements are $(Y_u^{(2)})_{ij}=y_u^{(2)}(U_L)^*_{3i} (U_R)_{3j}$, $(Y_d^{(2)})_{ij}=y_d^{(2)}(D_L)^*_{3i} (D_R)_{3j}$, hence only the last row of the unitary matrices is relevant, which we parametrize as:
\begin{eqnarray}
(U_L)_{31}&=&e^{i\rho_{u_L}}\sin\theta_{u_L}\sin\omega_{u_L}\;,\nonumber \\
(U_L)_{32}&=&e^{i\xi_{u_L}} \sin\theta_{u_L}\cos\omega_{u_L}\;,\nonumber\\
(U_L)_{33}&=&\cos\theta_{u_L}\;,
\end{eqnarray}
and similarly for $U_R$, $D_L$, $D_R$. In what follows we will neglect the phases for simplicity.

\section{Quantum effects on the quark masses and mixing angles}

We calculate now the impact of the quantum effects on the Yukawa couplings leading to fermion masses, $Y_u^{(1)}$ and $Y_d^{(1)}$. The one loop corrected couplings approximately read:
\begin{align}
&Y_u^{(1)}|_{\rm 1-loop} \simeq Y_u^{(1)}|_{\rm tree}+ \frac{1}{16\pi^2}\beta_u^{(1)}\log\frac{\Lambda}{M_H}\;,\nonumber\\ 
&Y_d^{(1)}|_{\rm 1-loop}\simeq Y_d^{(1)}|_{\rm tree}+ \frac{1}{16\pi^2}\beta_d^{(1)}\log\frac{\Lambda}{M_H}\;,
\label{eq:one-loop-corrected}
\end{align}
where $\Lambda$ is the cut-off scale of the theory and $\beta_u^{(1)}$, $\beta_d^{(1)}$ are the beta-functions,  which are included in the Appendix.

We find that quantum effects generate a rank-2 matrix, due to Feynman diagrams with the Higgs field $\Phi_2$ in the loop. The values of the Yukawa eigenvalues and the CKM matrix elements can be straightforwardly calculated from Eq.~(\ref{eq:one-loop-corrected}) using perturbation theory. Under the reasonable assumption $y_d^{(1)},y_d^{(2)}\ll y_u^{(1)},y_u^{(2)}$ (motivated by the empirical fact that $y_d^{(1)}\ll y_u^{(1)}$), the ratios between the Yukawa couplings of the second and third generation approximately read:
\begin{eqnarray}
\frac{y_c}{y_t}&\displaystyle{\simeq \left(\frac{1}{16\pi^2}\log\frac{\Lambda}{M_H}\right) \frac{3}{4} (y_u^{(2)})^2 \sin2\theta_{u_L}\sin2\theta_{u_R} }\;,\nonumber \\
\frac{y_s}{y_b}&\displaystyle{\simeq \left(\frac{1}{16\pi^2}\log\frac{\Lambda}{M_H}\right) \frac{y_u^{(1)} y_u^{(2)} y_d^{(2)}}{y_d^{(1)}}\cos\theta_{u_R}\sin\theta_{d_R} N_d}\;,
\label{eq:yukawa-ratios}
\end{eqnarray}
where 
\begin{align}
N_d&=\left[9 \sin^2\theta_{d_L}\cos^2\theta_{u_L}+4\cos^2\theta_{d_L}\sin^2\theta_{u_L}  -3 \sin2\theta_{d_L}\sin2\theta_{u_L}\cos(\omega_{d_L}-\omega_{u_L})\right]^{1/2}\;,
\label{eq:Nd}
\end{align}
which are loop suppressed but enhanced by the large logarithm of the cut-off scale over the heavy Higgs mass.  The dominant contribution to the charm quark mass is generated by a wave-function renormalization diagram proportional to $ \mathrm{Tr}( Y_u^{(1)}Y_u^{(2)\dagger})Y_u^{(2)}$, which requires a non-vanishing coupling of the Higgs $\Phi_2$ to the top quark as well as to the lighter generations of up-type quarks, which in turn imply, respectively, $\cos\theta_{u_L}\cos\theta_{u_R}\neq 0$ and $\sin\theta_{u_L}\sin\theta_{u_R}\neq 0$ in order to communicate the electroweak symmetry breaking from the third to the second generation. On the other hand, the dominant contribution to the strange quark mass is generated by a wave-function renormalization diagram proportional to $ \mathrm{Tr}( Y_u^{(2)}Y_u^{(1)\dagger}) Y_d^{(2)}$ and a vertex diagram proportional to $Y_u^{(2)}Y_u^{(1)\dagger}Y_d^{(2)}$. The former contribution requires, as above, a non-vanishing coupling of the Higgs $\Phi_2$ to the top quark as well as to the lighter generations of down-type quarks, which respectively imply $\cos\theta_{u_L}\cos\theta_{u_R}\neq 0$ and $\sin\theta_{d_L}\sin\theta_{d_R}\neq 0$, while the latter requires a non-vanishing coupling of the right-handed (left-handed) top quark to the lighter generations of left-handed (right-handed) quarks, which implies $\cos\theta_{u_R}\sin\theta_{u_L}\neq 0$ ($\cos\theta_{d_L}\sin\theta_{d_R}\neq 0$).

Notice that the first generation quarks remain massless in this simple scenario. They could be also generated radiatively if additional flavor structures were introduced in the model ({\it e.g.} by adding a third Higgs doublet or by postulating the existence of approximate rank-2 matrices at tree level).  We also note that the same result arises if the tree level Yukawa matrix is rank-2 but with Yukawa eigenvalues displaying very large hierarchies. If this is the case, the one loop contributions to the strange and charm masses induced by the third generation quarks will be much larger than the corresponding tree level values and, consequently, the masses at the one loop level will still be well approximated by Eq.(\ref{eq:yukawa-ratios}).

It is important to remark that the radiatively generated charm and strange masses depend logarithmically on the heavy Higgs mass, while flavor violating effects are suppressed by four powers of the latter. Therefore, by postulating a very large value for the heavy Higgs mass the predicted rates for the flavor violating processes will be within the experimental ranges.  More specifically, for arbitrary flavor structures, the measurement of the $K_L-K_S$ mass difference requires a heavy Higgs mass $M_H\gtrsim 150\,{\rm TeV}$~\cite{Shanker:1981mj}. While the direct production of the heavy states is far beyond the reach of present and foreseeable collider experiments, the new physics states produce deviations in flavor physics observables from the Standard Model values that might be at the reach of future experiments, depending on the value of the heavy Higgs mass.

The 12 and 21 elements of the CKM matrix are also calculable and read:
\begin{align}
V_{us}\simeq -V_{cd}\simeq \frac{3 \sin\theta_{d_L}\cos\theta_{u_L}\sin(\omega_{d_L}-\omega_{u_L})}{N_d}\;,
\label{eq:V12}
\end{align}
while the 11 and 22 elements are $V_{ud}\simeq V_{cs}\simeq \sqrt{1-V_{us}^2}$. Notably, the Cabibbo angle is not loop suppressed.  The reason lies in the ambiguity in the choice of the eigenvectors that diagonalize the tree level matrices $Y_u^{(1)}Y_u^{(1)\dagger}|_{\rm tree}$ and $Y_d^{(1)}Y_d^{(1)\dagger}|_{\rm tree}$ due to their two vanishing eigenvalues. When the perturbation is added, one non-vanishing eigenvalue is generated and the ambiguity is resolved, resulting in well defined eigenvectors which lead in turn to a well defined Cabibbo angle. In the perturbation theory language, the Cabibbo angle is generated at zero-th order.  In the renormalization group language, this effect can be interpreted as an infrared quasi-fixed point for the Cabibbo angle, that depends on the value of the corresponding beta-function, but is independent of the value of the Cabibbo angle at the cut-off scale. This behavior was noted in \cite{Ellis:1999my,Casas:1999tp} and extensively discussed in \cite{Casas:1999tg} for the mixing angles in the neutrino sector in the presence of degenerate mass eigenvalues.  Furthermore, the Cabibbo angle, in contrast to the radiatively generated masses, depends only on left-handed sector parameters. In particular, it is needed a misalignment between the left-handed up- and down-type quarks of the first two generations, namely $\sin(\omega_{d_L}-\omega_{u_L})\neq 0$, in order to generate a non-vanishing Cabibbo angle.

The remaining elements of the CKM matrix are:
\begin{align}
V_{ub}&\simeq \left(\frac{1}{16\pi^2}\log\frac{\Lambda}{M_H}\right) \frac{3 y_u^{(1)} y_u^{(2)} y_d^{(2)}}{y_d^{(1)}}\sin\theta_{d_L}\cos\theta_{d_R}\cos\theta_{u_L}\cos\theta_{u_R}\sin(\omega_{d_L}-\omega_{u_L}) \;, \nonumber \\
V_{cb}&\simeq \left(\frac{1}{16\pi^2}\log\frac{\Lambda}{M_H}\right) 
\frac{ y_u^{(1)} y_u^{(2)} y_d^{(2)}  }{y_d^{(1)}} 
 \left\{ \frac{1}{4}\frac{y_d^{(1)} y_u^{(2)}}{y_d^{(2)} y_u^{(1)}} \sin 2 \theta _{u_L}(3 \cos 2 \theta _{u_R}+2)\right. \nonumber\\ 
& +\left. \cos \theta _{d_R} \cos \theta _{u_R} \left[2 \cos \theta _{d_L} \sin \theta _{u_L}-3 \sin \theta _{d_L} \cos \theta _{u_L} \cos (\omega _{d_L}-\omega _{u_L})\right]\right\} \;,
\label{eq:V23}
\end{align}
while $V_{td}=-V_{ub}V_{cs}+V_{us}V_{cb}$ and $V_{ts}=-V_{cb}V_{ud}+V_{ub}V_{cd}$, as required by unitarity, and $V_{tb}\simeq 1$. In contrast to the Cabibbo angle, all other off-diagonal entries of the CKM matrix are generated at first order of perturbation theory and are therefore expected to be much smaller than the 12 entry, in qualitative agreement with experiments. Moreover, these elements depend on right-handed sector parameters, similarly to the radiatively generated quark masses.

The measured values of $y_c/y_t$, $y_s/y_b$ and the CKM matrix can be accommodated within this framework by choosing appropriate model parameters. We note first that the right-handed angles $\theta_{d_R}$ and $\theta_{u_R}$ are univocally determined by the  quark parameters:
\begin{align}
&\frac{y_s}{y_b}\frac{V_{us}}{V_{ub}} \simeq 
\tan\theta_{d_R}\;, \nonumber \\
&\frac{y_c}{y_t}\frac{V_{us}}{V_{td}}\simeq 
\frac{3\sin2\theta_{u_R}}{2+3\cos2\theta_{u_R}}\;,
\end{align}
 which approximately give $\theta_{u_R}\approx 0.16$, $\theta_{d_R}\approx 1.06$. On the other hand, there are degeneracies among the remaining parameters. One possible choice is $y_u^{(2)}\approx 1.04$, $y_d^{(2)}\approx 0.02$, $\theta_{d_L}\approx 0.61$, $\theta_{u_L}\approx 0.51$, $\omega_{d_L}-\omega_{u_L}\approx 0.10$. It is notable that under the reasonable assumptions that the coupling $y_u^{(2)}$ ($y_d^{(2)}$) is of the same order as $y_u^{(1)}$ ($y_d^{(1)}$) and that the mixing angles are all ${\cal O}(0.1)$ it is possible to naturally reproduce the measured masses of the second generation quarks and the mixing angles.  A similar scheme could be responsible for the charged lepton masses in the presence of right-handed neutrinos, due to the quark-lepton symmetry in the type I see-saw mechanism. The implications for the neutrino masses and mixing angles will be discussed elsewhere \cite{ISB}.

The framework presented here contains a large number of free parameters and does not lead to any prediction. Nevertheless, the degeneracies could be broken by incorporating to the analysis other flavor observables, such as deviations from the Standard Model predictions in flavor changing neutral currents, which could be measured in future experiments.

\section{Conclusions}

The hierarchies among the quark masses of different generations, as well as the hierarchies among the quark mixing angles, strongly suggest the existence of a dynamical mechanism to generate this pattern. We have argued that a second Higgs doublet added to the Standard Model particle content, with no additional fermions nor additional symmetries, can be responsible for generating via quantum effects a mass hierarchy between the second and third quark generations and a pattern of mixing angles in qualitative agreement with observations. This scheme can reproduce the measured values even in the decoupling limit of the heavy Higgs, therefore the strong constraints on a second Higgs doublet from flavour changing neutral currents  can be easily avoided if  the heavy Higgs mass is sufficiently large. On the other hand, if the new physics scale is low enough, new phenomena could be observed in experiments at the intensity frontier, opening the possibility to test this mechanism.

\section*{Acknowledgements}

We are grateful to Camilo Garcia-Cely for useful discussions. This work was supported in part by the DFG cluster of excellence ``Origin and Structure of the Universe'' and by the ERC Advanced Grant project ``FLAVOUR''(267104) (A.I.).

\section*{Note Added}
While this work was being finalized, we learned of the work \cite{AFH}, where it is presented a supersymmetric framework to  radiatively generate quark masses and mixing angles.

\appendix
\section{Appendix: beta functions}

The Renormalization Group Equations of the quark Yukawa couplings $Y_{u,d}^{(a)}$ can be cast as:
\begin{equation}
16\pi^2\frac{d Y_u^{(a)}}{d\log\mu}=\beta_u^{(a)}, ~~~~
16\pi^2\frac{d Y_d^{(a)}}{d\log\mu}=\beta_d^{(a)},
\end{equation}
where the beta-functions were calculated in \cite{Grimus:2004yh,Cheng:1973nv,Cvetic:1997zd} and are given by:
\begin{eqnarray}
\beta_u^{(a)} &=&
\left( - 8 g_s^2 - \frac{9}{4}\, g^2 - \frac{17}{12}\, {g^\prime}^2\right) Y_u^{(a)} + \sum_{b=1,2} T_{ab}^* Y_u^{(b)}
\nonumber
\\ 
& +& \sum_{b=1,2} \left(
- 2\, Y_d^{(b)} Y_d^{(a)\dagger} Y_u^{(b)}
+ \frac{1}{2} Y_d^{(b)} Y_d^{(b)\dagger} Y_u^{(a)}
+ Y_u^{(a)} Y_u^{(b)\dagger} Y_u^{(b)}
+ \frac{1}{2}Y_u^{(b)} Y_u^{(b)\dagger} Y_u^{(a)}
\right), 
\nonumber \\
\beta_d^{(a)} &=&
\left(- 8 g_s^2 - \frac{9}{4}\, g^2 - \frac{5}{12}\, {g^\prime}^2\right) 
Y_d^{(a)} + \sum_{b=1,2} T_{ab} Y_d^{(b)}
\nonumber
\label{uvhfw2}
\\ 
 &+& \sum_{b=1,2} \left(
- 2\, Y_u^{(b)} Y_u^{(a)\dagger} Y_d^{(b)}
+ \frac{1}{2} Y_u^{(b)} Y_u^{(b)\dagger} Y_d^{(a)}
+ Y_d^{(a)} Y_d^{(b)\dagger} Y_d^{(b)}
+ \frac{1}{2} Y_d^{(b)} Y_d^{(b)\dagger} Y_d^{(a)}
\right).
\nonumber \\
\end{eqnarray}
Here $g_s$, $g$ and $g'$ the $SU(3)_C$, $SU(2)_L$ and $U(1)_Y$ gauge coupling constants, respectively, and 
\begin{align}
T_{ab} ={\rm Tr} \left( 3 Y_d^{(a)} Y_d^{(b)\dagger}
+ 3 Y_u^{(a)^\dagger} Y_u^{(b)}+
Y_e^{(a)} Y_e^{(b)\dagger }\right).
\end{align}

\end{document}